\documentclass[a4paper,12pt]{article}
\usepackage[cp1251]{inputenc}
\usepackage[russian]{babel}
\usepackage{latexsym}
\usepackage{amsmath,amssymb}
\begin{document}

\noindent
 Universal Decimal Classification
\\
\\
{\bf ASYMMETRICAL PSEUDOELASTICITY}
 \bigskip
\\
{\bf V.O.~Bytev}, Tyumen State University\\
 Tyumen, 625003, vbytev@utmn.ru
\\
{\bf L.I.~Shkutin}, Institute of Computer Modeling\\
 Siberian Division, Russian Academy of Sciences\\
 Krasnoyarsk, 660036, shkutin@icm.krasn.ru
 \bigskip
 \\

Term ''asymmetrical pseudoelasticity'' refers to the theory, in
which a symmetrical stress tensor and a symmetrical strain tensor
are connected by means of an asymmetrical material tensor. An
6-dimensional asymmetrical matrix of elasticity has been
constructed that is invariable in relation to orthogonal
transformation with a single rotation operator and coordinated
with conservation laws of the continuum mechanics. The matrix has
got eight independent components and expands the traditional
definition of transversally isotropic (hexagonal) material
symmetry. The suggested theory includes definition of
3-dimensional and 2-dimensional linear boundary value problems and
accurate solutions generalizing the traditional polynomial
solutions, A.Love's solutions, and N.I.Muskhelishvili's solutions
and providing new kinematic effects.


{\bf Introduction}. The problem of detecting the constitutive relations
within the continuum mechanics was set in the works of Truesdell's school
[1] and the idea of applying continuous groups theory for its solution
was ``in the air'' as back in 1912-1914 A.Einstein pointed out the
necessity to look for invariants of continuous transformations groups as
one of major problems of the physics. The modern stage of systematic
application of group analysis methods for models of the continuum
mechanics was developed in the works of L.V.Ovsyannikov's [2] and
N.Kh.Ibragimov's schools. Group features of the differential equations of
flow media and gas media mechanics are studied in detail at present.

Invariance principle forms the basis for modern approaches connected with
searching for new types of physical structures. Within the continuum
mechanics it provides the general approach to composing defining
relations that are required for closure of differential equation sets
resulting from integral conservation laws.

The method of solving problems of synthesis and analysis of constitutive
equations within the continuum mechanics basing on the invariance
principle was suggested in O.V.Bytev's works [4-6]. Differential equation
set of purely mechanical non-polar continuum
\[
{\bf v}_t+({\bf v}\cdot\nabla){\bf v}- \rho^{-1} \mathrm{div}\,
{\bf T}+ \rho^{-1}\nabla p= {\bf 0}\,,
\]
\[
p_t+ ({\bf v}\cdot\nabla)p+ G\, \mathrm{div}\, {\bf v}+ H\, {\bf
T} : \nabla {\bf v}=0\,,
\]
\[
\rho_t+ \mathrm{div}\, (\rho {\bf v})=0\,, \ G=G(p,\rho)\,, \
H=H(p,\rho)\,, \ {\bf T}= {\bf T}(\nabla {\bf v})\,,
\]
resulting from conservation laws and traditional two-parameter
thermodynamics were subject to group analysis with the purpose of
using a group of continuous transformations for formation of the
invariance principle itself (symbols used in the equations: {\bf
v} is the vector velocity field, {\bf T} is the tensor field of
dissipative stress, $p$ is the balance pressure, $\rho $ is the
continuum mass density, $G$ and $H$ are arbitrary functions of
state parameters).

Group analysis performed in [5, 6] detected that transformation of
initial equations equivalency does not include $SO_{3}$ group. Thus, a
full classification problem was solved without any additional suggestions
(isotropy, coaxiality etc). The procedure of group classification
detected the following possibilities: 1) if {\bf T}, $G$ and $H$ are
arbitrary functions of their arguments, the initial system allows only
Hamilton's group that forms the group germ; 2) on assumption of three
rotations (spherical invariance), there are traditional dependencies
between {\bf T }and $\nabla{\bf v}$; 3) on assumption of one rotation
(rotational invariance), dependencies between the continuum state
parameters can differ from traditional ones.

The hierarchy of constitutive relations synthesized in [6] allows to
formulate non-traditional closed models of liquid and solid media. Thus,
analysis of the linear dependency between a symmetrical tensor of viscous
stresses and a symmetrical tensor of strain rates in case when a medium
model assumes only one rotation operator proved that these tensors can be
connected by an asymmetrical transformation tensor.

On the basis of results obtained in [5, 6], some comments shall be made in
respect of the traditional approach to formation of models within the
continuum mechanics.

1. Theoretical approach (that has already become traditional) used
to determine the stress tensor dependency of strain tensor (of
strain rates tensor) is based on a postulate of stress tensor
invariance in relation to actions of $SO_{3}$ group [1]. It
results in symmetrical models of continuous media being in
subordination to coaxiality of adjoint tensors, whereas, for
instance, in ground models it is impossible to preserve coaxiality
of stress and strain rate tensors. It is thus worth mentioning
that a stress tensor is not an abstract tensor object, it complies
with the momentum conservation law. Here the question arises: is
the momentum conservation law always invariant in relation to
$SO_{3}$ group action?

2. Bringing two tensor fields of the second rank to linear non-vector
proportionality is equivalent to bringing two quadratic forms to a canonical
form by means of a single orthogonal transformation. It is possible only if
a pair of nondegenerate quadratic forms uses a regular pencil, i.e. when
both coefficient matrices are symmetrical and one of the associated
quadratic forms is positively definite (Sylvester's criterion) [7]. Stress
tensors and deformation tensors are symmetric by definition of a simple
non-polar continuum, whereas the Sylvester's criterion shall be checked.

3. The mechanics of a deformed solid body distinguishes the Cauchy
elasticity when a stress tensor is defined as an invariant function of a
deformation tensor and Green elasticity when elastic energy potential is
postulated. According to C.Truesdell, a material with Green elasticity is
referred to as hyperelastic. The procedure of composing defining equations
with application of elastic potential being the quadratic form of
deformation tensor components that is usual for invariant hyperelasticity
theory excludes the asymmetry of elasticity tensor.

The foregoing comments formed the basis for a more detailed analysis of
possible variants of defining relations within Cauchy's linear elasticity
theory.
  \vskip 4mm

 \setcounter{equation}{0}
 \def\theequation{1.\arabic{equation}}
{\bf 1.3-Dimensional Model of the Asymmetrical Pseudoelasticity}

  \vskip 4mm

Let us analyze a 3-dimensional linear model of elastic medium with
asymmetric elasticity tensor (of rigidity or compliance) in the Cartesian
reference system $x_1,\,x_2,\,x_3$. Let $\sigma_{ij}$ be components of a
symmetrical stress tensor, $\varepsilon_{ij}$ -- components of a
symmetrical deformation tensor ($i,j = 1,\, 2,\, 3$). Let us introduce
the following symbols:
\[
\begin{array}{l}
 \sigma _1 = \sigma _{11}\,,\,\, \sigma _2 = \sigma _{22}\,,\,\,
 \sigma _3 = \sigma _{12} = \sigma _{21} \,, \\
 \sigma _4 = \sigma _{31} = \sigma _{13} \,,\,\, \sigma _5 =
\sigma_{32} = \sigma _{23} \,,\,\, \sigma _6 = \sigma _{33} \,, \\
\varepsilon _1 = \varepsilon _{11} \,,\,\,\varepsilon _2 = \varepsilon
_{22} \,,\,\,\varepsilon _3 = 2\,\varepsilon _{12} = 2\varepsilon _{21}
\,,\\  \varepsilon _4 = 2\varepsilon _{31} = 2\varepsilon _{13} \,,\,\,
\varepsilon _5 = 2\varepsilon _{32} =
2\varepsilon _{23} \,,\,\,  \varepsilon _6 = \varepsilon _{33} \\
 \end{array}
\]
and column-vectors
\[
\sigma = \,\left[ {\sigma _1 \,,\,\,\sigma _2 \,,\,\,\sigma _3
\,,\,\,\sigma _4 \,,\,\,\sigma _5 \,,\,\,\sigma _6 }
\right]^T,\,\,\,\,\,\varepsilon = \,\left[ {\varepsilon _1
\,,\,\,\varepsilon _2 \,,\,\,\varepsilon _3 \,,\,\,\varepsilon _4
\,,\,\,\varepsilon _5 \,,\,\,\varepsilon _6 } \right]^T.
\]

The most general linear connection between Cartesian tensors
$\sigma_{ij}$ and $\varepsilon _{ij}$ is expressed by the
constitutive relation $\sigma_M = C_{MN}\varepsilon_N$, where
capital indices range over values 1~--~6, thus $C_{MN}$
coefficients form the elasticity matrix of the $6^{th}$ rank. Let
us perform an orthogonal transformation with a single rotation
operator with the analyzed elasticity relation and define
conditions of this relation being invariable in relation to this
transformation. In order to do so, we shall perform a right-hand
rotation of axes in relation to $x_{3}$ axis:
\[
 x'_1 = x_1\cos\varphi + x_2\sin\varphi, \qquad
 x'_2 =-x_1\sin\varphi + x_2\cos\varphi, \qquad x'_3 = x_3\,,
\]
where $\varphi$ is a random angle. Strain tensor (and stress tensor)
components are transformed according to the rule:
\[
\begin{array}{l}
 \varepsilon'_{11} = \varepsilon_{11}\cos^2\varphi + \varepsilon_{22}
 \sin^2\varphi + 2\varepsilon_{12}\cos\varphi\sin\varphi\,, \\
 \varepsilon'_{22} = \varepsilon_{11}\sin^2\varphi + \varepsilon_{22}
 \cos^2\varphi - 2\varepsilon_{12}\cos\varphi\sin\varphi\,, \\
 \varepsilon'_{12} =\varepsilon_{12}(\cos^2\varphi - \sin^2\varphi) -
 (\varepsilon_{11}-\varepsilon_{22})\cos\varphi\sin\varphi\,, \\
 \varepsilon'_{31} =\ \ \varepsilon_{31}\cos\varphi + \varepsilon_{32}
 \sin\varphi\,, \\
 \varepsilon'_{32} =  - \varepsilon_{31}\sin\varphi + \varepsilon_{32}
 \cos\varphi\,, \ \  \varepsilon'_{33} = \varepsilon_{33}\,.
  \end{array}
\]

The condition of rotational invariance of elasticity relations is
expressed by formula $\sigma'_M = C_{MN}\varepsilon'_N$\ considering the
rule of tensors transformation. Thus, relation $\sigma'_6 = C_{6N}
\varepsilon'_N$\ requires the following equalities being fulfilled
\[
 C_{62}=C_{61}\,, \ \ C_{63}=C_{64}=C_{65}=0\,,
\]
relation $\sigma'_4 = C_{4N}\varepsilon'_N$\ requires fulfillment of
equalities
\[
\begin{array}{l}
 C_{41}=C_{42}=C_{43}= C_{46}=0\,, \\
 C_{51}=C_{52}=C_{53}= C_{56}=0\,, \ \ C_{54}=-C_{45}\,, \ \
 C_{55}=C_{44}
  \end{array}
\]
and relation $\sigma'_2 = C_{2N}\varepsilon'_N$\ requires fulfillment of
equalities
\[
\begin{array}{l}
 C_{14}=C_{15}=C_{24}= C_{25}=C_{34}=C_{35}=C_{36}=0\,, \\
 C_{21}=C_{12}\,, \ \ C_{22}= C_{11}\,, \ \ C_{23}=-C_{13}\,, \ \
 C_{26}=C_{16}\,,  \\ C_{31}=-C_{13}\,, \ \ C_{32}= C_{13}\,, \ \
2C_{33}= C_{11}-C_{12}\,.
  \end{array}
\]
The other of equations do not provide any additional restrictions.

Ultimately, the system of constitutive relations for a linear model with a
single rotation operator can be represented in a matrix form
\begin{equation}
\sigma = C\,\varepsilon\,, \ \ C = \left[ {{\begin{array}{*{20}c}
\,\,\,\,{C_{11} } \hfill & \,\,{C_{12} } \hfill & \,\,{C_{13} } \hfill &
\,\,\,0 \hfill & \,\,0 \hfill & \,{C_{16}} \hfill
 \\
\,\,\,\,{C_{12} } \hfill & \,\,{C_{11} } \hfill & \!\!{ - C_{13} } \hfill
& \,\,\,0 \hfill & \,\,0 \hfill & \,{C_{16}} \hfill
 \\
{ - C_{13} } \hfill & \,\,{C_{13} } \hfill & \,\,{C_{33} } \hfill &
\,\,\,0 \hfill & \,\,0 \hfill & \,\,0 \hfill
 \\
\,\,\,\,\,\,\,0 \hfill & \,\,\,\,0 \hfill & \,\,\,\,0 \hfill &
\,\,{C_{44} } \hfill & \,{C_{45} } \hfill & \,\,0 \hfill
 \\
\,\,\,\,\,\,\,0 \hfill & \,\,\,\,0 \hfill & \,\,\,\,0 \hfill & \!\! { -
C_{45} } \hfill & \,{C_{44} } \hfill & \,\,0 \hfill
 \\
\,\,\,\,{C_{61} } \hfill & \,\,\,\,{C_{61} } \hfill & \,\,\,\,0 \hfill &
\,\,\,0 \hfill & \,\,0 \hfill & \,{C_{66} } \hfill \\
\end{array} }} \right]\,,
\end{equation}
where $C$ is elasticity matrix $6\times 6$. There is equality
$C_{12} = C_{11} \!- 2 C_{33}$, the remaining 8 components:
$C_{11}$, $C_{13}$, $C_{16}$, $C_{33}$, $C_{44}$, $C_{45}$,
$C_{61}$, $C_{66}$ are independent kinetic parameters, moreover
C$_{13}$ and C$_{45}$ can be either positive or negative.

Elasticity matrix present in (1.1) is different from a traditional matrix
of material rigidness with transversally isotropic (hexagonal) symmetry
by presence of asymmetric components and transforms into traditional one
on conditions $C_{45}=C_{13}=0$, $C_{61}=C_{16}$\,.

Sylvester's criterion confirms that relation (1.1) ensures
positive definiteness of a dissipative function. However, unlike
the traditional elasticity, the dissipative function does not have
a potential. The fact of an elasticity matrix having asymmetric
components means that stress tensor and deformation tensor are not
coaxial. The indicated peculiarities drastically differ the
asymmetric elasticity model from the symmetrical one and impose
the authors to introduce a new term --- ''pseudoelasticity'' ---
for it.

The relation (1.1) results in the following inverse dependence
\begin{equation}
\varepsilon = D\,\sigma ,\,\,\,\,D = C^{ - 1}.
\end{equation}
Compliance matrix $D$ has the same structure as $C$. Its zero components
are
\[
\begin{array}{l}
 D_{14} = D_{15} = D_{24} = D_{25} = D_{34} = D_{35} = D_{36} =
 D_{41} = D_{42} = \\
 D_{43} = D_{46} = D_{51} = D_{52} = D_{53} = D_{56} =
 D_{63} = D_{64} = D_{65} = 0\,.
\end{array}
\]
Therefore relation (1.2) can be expressed in the form of a system
\begin{eqnarray}
\varepsilon _1 = D_{11} \sigma _1 + D_{12} \sigma _2 + D_{13}
 \sigma _3 + D_{16} \sigma _6 \,,\,\,
\varepsilon _2 = D_{21} \sigma _1 + D_{22} \sigma _2 + D_{23}
 \sigma _3 + D_{26} \sigma _6 \,, \nonumber \\
\varepsilon _3 = D_{31} \sigma _1 + D_{32} \sigma _2 + D_{33}
 \sigma _3 \,,\,\,
\varepsilon _4 = D_{44} \sigma _4 + D_{45} \sigma _5 \,, \\
\varepsilon _5 = D_{54} \sigma _4 + D_{55} \sigma _5 \,,\,\, \varepsilon
_6 = D_{61} \sigma _1 + D_{62} \sigma _2
 + D_{66} \sigma_6 \,. \nonumber
\end{eqnarray}

Dynamic and kinematic equations preserve the traditional form. Within the
Cartesian reference system the dynamic equations have the following form
\begin{equation}
\sigma _{1,1} + \sigma _{3,\,2} + \sigma _{4,\,3} + F_1 = 0\,, \,\,
\sigma _{3,1} + \sigma _{2,\,2} + \sigma _{5,\,3} + F_2 = 0\,, \,\,
\sigma _{4,1} + \sigma _{5,\,2} + \sigma _{6,\,3} + F_3 = 0\,, \,\,
\end{equation}
where $F_{i}$ means components of volume forces (including inertial
forces), index after a comma means a derivative at the corresponding
coordinate.

If $w_i(x_1,\,x_2,\,x_3)$ are displacements of the body points, then
traditional kinematic equations are
\begin{equation}
\varepsilon _1 = w_{1,1}\,, \,\, \varepsilon _2 = w_{2,\,2}\,, \,\,
\varepsilon _3 = w_{1,\,2} + w_{2,1}\,, \,\, \varepsilon _4 = w_{1,\,3} +
w_{3,1}\,,\,\,\varepsilon _5 = w_{2,\,3} + w_{3,\,2}\,, \,\, \varepsilon
_6 = w_{3,\,3}\,.
\end{equation}
Resolvability of this system in relation to $w_{i}$ functions is
provided for by the traditional equations of strains compatibility
[8--10].

Equations (1.3)~--~(1.5) form a closed set within an 3-dimensional
asymmetric model of Cauchy elastic continuum.
  \vskip 2mm

{\bf Accurate Solutions of 3-Dimensional Problems.} The solvability
conditions of system (1.5) are fulfilled identically for deformations
being in linear dependence of the coordinates. Similar to the traditional
elasticity theory [9], a class of problems having an analytical solutions
with a linear field of strains may be distinguished (separated???).

Set of equations (1.3)~--~(1.5) shall be analyzed with the
following assumptions: 1) components of matrices C and D do not
depend on the coordinates; 2) components of $F_{i }$ external
forces do not depend on the coordinates and time. According to the
first assumption, the object of the analysis is a homogeneous body
with an asymmetric elasticity tensor. Because of the indicated
restrictions, the following linear functions are the solutions of
equations (1.4):
\begin{equation}
\sigma _j = a_j\, x + b_j\, y + c_j\, z + p_j
\end{equation}
($x,\,y,\,z$ are Cartesian coordinates) with coefficients $a_j\,,\,
b_j\,,\, c_j\,,\, p_j$\,, connected with three equalities
\[
c_4 = - a_1 - b_3 - F_1 \,,\,\, c_5 = - a_3 - b_2 - F_2 \,,\,\, c_6 = -
a_4 - b_5 - F_3 \,.
\]
As a result of (1.3), the strains shall also be linear functions of
coordinates
\begin{equation}
 \varepsilon _k = f_k\, x + g_k\, y + h_k\, z + e_k\,,
\end{equation}
with coefficients $e_k\,,\, f_k\,,\, g_k\,,\, h_k$\,, calculated via
$a_j\,,\, b_j\,,\, c_j\,,\, p_j$\,.

The conditions of strains compatibility identically allow a quadratic
dependence of displacements of coordinates [9]:
\begin{equation}
\begin{array}{l}
 w_i = {\textstyle{1 \over 2}} \left( {\alpha_{i1}\,x^2 +
 \alpha_{i2}\,y^2 + \alpha_{i3}\,z^2} \right) \ + \\
 \alpha_{i4}\, xy + \alpha_{i5}\,xz + \alpha_{i6}\,yz +
 \alpha_i\,x + \beta_i\,y + \gamma_i\,z + \delta_i \,. \\
\end{array}
\end{equation}
These functions contain 30 unknown constants \textit{$\alpha
$}$_{ij}$\textit {, $\alpha $}$_{i}$\textit{, $\beta $}$_{i}$\textit{,
$\gamma $}$_{i}$\textit{, $\delta $}$_{i}$. Their number can be reduced
if to eliminate the solid body displacements.

Exclusion of displacements and rotations of the body in point
(0,~0,~0) by means of conditions
\[
w_i = 0\,,\,\,\, w_{1,\,3} = 0\,,\,\,\, w_{2,\,3} = 0\,,\,\,\, w_{2,1} -
w_{1,\,2} = 0
\]
brings us to equalities
\[
\delta _i = 0\,,\,\, \gamma _1 = 0\,,\,\, \gamma _2 = 0\,,\,\, \alpha _2
- \beta _1 = 0\,,
\]
that eliminate six constants. The remaining coefficients of polynomial
(1.8) are defined from (1.5) by putting the known functions (1.7) into
their left parts and the unknown functions (1.8) into the right parts.
Thus, all coefficients of polynomial (1.8) can be defined:
\[
\begin{array}{l}
\alpha _{11} = f_1 \,,\,\,\alpha _{12} = g_3 - f_2 \,,\,\,\alpha _{13} =
h_4 - f_6 \,,
\\
\alpha _{14} = g_1 \,,\,\,\alpha _{15} = h_1 \,,\,\,2\alpha _{16} = h_3 +
g_4 - f_5 \,,
\\
\alpha _{21} = f_3 - g_1 \,,\,\,\alpha _{22} = g_2 \,,\alpha _{23} = h_5
- g_6 \,,
\\
\alpha _{24} = f_2 \,,\,\,2\alpha _{25} = f_5 + h_3 - g_4 \,,\,\,\alpha
_{26} = h_2 \,,
\\
\alpha _{31} = f_4 - h_1 \,,\,\,\alpha _{32} = g_5 - h_2 \,,\,\,\alpha
_{33} = h_6 \,,
\\
2\alpha _{34} = f_5 - h_3 + g_4 \,,\,\,\alpha _{35} = f_6 \,,\,\,\alpha
_{36} = g_6 \,,
\\
\alpha _1 = e_1 \,,\,\,2\alpha _2 = 2\beta _1 = e_3 \,,\,\,\beta _2 = e_2
\,,\,\,\alpha _3 = e_4 \,,
\\
\beta _3 = e_5 \,,\,\,\gamma _1 = \gamma _2 = 0\,,\,\,\gamma _3 = e_6
\,,\,\,\delta _i = 0\,.
\end{array}
\]

Functions (1.6)~--~(1.8) give an accurate polynomial solution of
the class of 3-dimensional problems on deformation of
asymmetrically-elastic bodies under volume and surface loads. Let
us analyze two typical boundary value problems from this class.
  \vskip 2mm

{\bf Circular Shaft Torsion.} Let $L$ be the shaft length, $R$ is
the cross-section radius. Let us match z axis (of material
rotational symmetry) with the shaft axis and locate the origin of
coordinates at the midsection, so that\, $\left( {x,y} \right)
\!\in \left[ {0,R} \right]$,\, $z \!\in \left[ { - {\textstyle{1
\over 2}} \,L\,, \,\,{\textstyle{1 \over 2}}\,L }\, \right]$.

Let us assume $F_{3}=F_{2}=F_{1}=0$ in equations (1.4) and analyze the
stress state
\[
\sigma_4 = b_4 y,\,\,\sigma_5 = a_5 x\,,\,\,\sigma_6 = \sigma_3 =
\sigma_2 = \sigma_1 = 0\,,
\]
which is the solution of homogeneous dynamic equations.

When analyzing the shaft torsion, let us request absence of tangential
stress in axial section. This condition is expressed by equalities
\[
\sigma _{zr} = \sigma _4 \cos \varphi + \sigma _5 \sin \varphi
 = b_4 r\sin \varphi \cos \varphi + a_5 r\cos \varphi \sin \varphi = 0\,.
\]
and shall be fulfilled provided that $b_{4}=-a_{5}$.

Thus the stress state
\begin{equation}
\sigma _4 = - a_5 y,\,\,\sigma _5 = a_5 x\,,\,\,\sigma _6 = \sigma _3 =
\sigma _2 = \sigma _1 = 0
\end{equation}
satisfies homogeneous equations of equilibrium and homogeneous boundary
conditions on the cylindric surface and is the state of the shaft pure
torsion.

Calculating the tangential stress in the shaft cross-section, we shall
have
\[
\sigma _{z\varphi } = \sigma _5 \cos \varphi - \sigma _4 \sin \varphi =
a_5 r.
\]
The obtained solution is accurate when the shaft is twisted by
tangential stress $\sigma _{z\varphi} = \tau r/R$, applied to ends
of the shaft. The constant $a_5 = \tau/R$ is defined by means of
$\tau$ parameter which has meaning of tangential stress at the
boundary contour of the shaft.

Strains corresponding to stresses (1.9) take the following values
\[
\varepsilon _4 = a_5 \left( {D_{45} x - D_{44} y} \right), \,\,\,
\varepsilon _5 = a_5 \left( {D_{44} x + D_{45} y} \right), \,\,\,
\varepsilon _6 =\varepsilon _3 =\varepsilon _2 =\varepsilon _1 = 0 \,,
\]
displacements are calculated by formulas
\[
w_1 = - D_{44} a_5 \,y\,z\,,\,\,\, w_2 = D_{44} a_5 \,x\,z\,,\,\,\, 2w_3
=  D_{45} a_5 \left( {x^2 + y^2} \right)
\]
in the Cartesian reference system or by formulas
\begin{equation}
w_r = 0\,,\,\,\,  w_\varphi = D_{44} a_5 \,r\,z\,,\,\,\,  2w_3 = D_{45}
a_5 \,r^2
\end{equation}
in the cylindrical system. The traditional solution results from (1.10)
when $D_{45} = 0$ [9]:
\begin{equation}
w_r = 0\,,\,\,\,w_\varphi = D_{44} a_5 \,r\,z\,,\,\,\,w_3 = 0\,.
\end{equation}
Product $D_{44} a_5 = \omega $ has the meaning of torsion angle per a
unit of the shaft length.

Comparison of formulas (1.10) and (1.11) shows that the solution of the
modified problem gives the displacements field being different from the
traditional one. It contains axial displacements variable along the
radius that generate deplanetion of the shaft cross sections (Fig.1).
  \vskip 4mm

{\bf Biaxial Bending of a Plate}. The problem is set for a rectangular
plate of constant thickness $2h$. The $x$ and $y$ axes are located within
the middle plane of the plate, $z$ axis -- along the central normal to
it. The coordinates are given in domain $x \!\in [ {-a,\,a} ]$, $y \!\in
[ {-b,\,b} ]$, $z \!\in [ {-h,\,h} ]$\, ($2a$, $2b$ are dimensions of the
plate in middle plane).

The plate is bending by stresses
\[
\sigma _1 = c_1 z\,,\,\,\, \sigma _2 = c_2 z\,,\,\,\, \sigma _3 = \sigma
_4 = \sigma _5 = \sigma _6 = 0\,.
\]
This stresses state satisfy the homogeneous equations of equilibrium and
homogeneous force conditions at surfaces $z=\mp h$. In solution of (1.6)
we have
\[
a_k = b_k = p_k = 0\,,\,\,\,c_6 = c_5 = c_4 = c_3 = 0\,.
\]
Non-zero deformations
\[
\varepsilon _1 = D_{11} \sigma _1 + D_{12} \sigma _2 \,,\,\,\,
\varepsilon _2 = D_{21} \sigma _1 + D_{22} \sigma _2 \,,\,\,\,
\varepsilon _3 = D_{31} \sigma _1 + D_{32} \sigma _2
\]
are linear functions of $z$ coordinate. Displacements are calculated by
formulae
\begin{equation}
w_1 = z\left( {\alpha _{15} x + \alpha _{16} y} \right)\,,\,\,\, w_2 =
z\left( {\alpha _{25} x + \alpha _{26} y} \right)\,,\,\,\, 2w_3 = \alpha
_{31}\, x^2 + \alpha _{32}\, y^2 + 2\alpha _{34}\, xy
\end{equation}
with constant coefficients (in case of given $c_{1}$ and $c_{2}$):
\[
\begin{array}{l}
 \alpha _{15} = D_{11} c_1 + D_{12} c_2 \,,\,\,
2\alpha _{16} = D_{31} c_1 + D_{32} c_2 \,,\,\,
 \alpha _{26} = D_{21} c_1 + D_{22} c_2 \,,\,\, \\
 \alpha _{25} = \alpha _{16}\,,\,\,\alpha _{31} = -\alpha _{15}\,,\,\,
 \alpha _{32} = -\alpha _{26}\,,\,\,\alpha _{34} = -\alpha _{16}\,.
\end{array}
\]

In case of an uniaxial bend by stress $\sigma _{2}$ ($\sigma _{1}=0$),
deformations of the plate sections corresponding to solution of (1.12)
are schematically represented in Fig.4-6. Unlike the traditional
solution, the plate is curved not only in relation to $y$ axis (Fig.4),
but in relation to $x$ axis as well (Fig.5). Fig.6 demonstrates level
lines on equidistant surfaces of the plate. We can notice the rotation of
asymptotes of hyperbolic lines, which is absent within the traditional
solution [9].
  \vskip 4mm

 \setcounter{equation}{0}
 \def\theequation{2.\arabic{equation}}
{\bf 2. 2-Dimensional Model of the Asymmetrical Pseudoelasticity}
  \vskip 4mm

If deformation is parallel to plane $Оxy$ (plane deformation), so that
$\varepsilon _{4}=\varepsilon _{5}=\varepsilon _{6}= 0$, a six-equations
system (1.1) degenerates into a three-equations system
\begin{equation}
 \sigma = A\,\varepsilon\,, \ \ A = \left[ {{\begin{array}{*{20}c}
 \,\,\,\,{C_{11} } \hfill & \,\,{C_{12} } \hfill &   \,\,{C_{13} }
 \hfill  \\
 \,\,\,\,{C_{12} } \hfill & \,\,{C_{11} } \hfill & \!\!{- C_{13} }
 \hfill \\
        {-C_{13} } \hfill & \,\,{C_{13} } \hfill &   \,\,{C_{33} }
 \hfill \\
\end{array} }} \right]\,.
\end{equation}
In other words, matrix $C$ degenerates into matrix $A$ of plane
deformation problem.

Let us introduce displacements vector $w=(u\,,\,v)$ and components of the
linear strains tensor
\[
 \varepsilon_{11}=\varepsilon_{1}= u_{,\,x}\,, \,\,\,\, \varepsilon_{22}
=\varepsilon_{2} =v_{,\,y}\,, \,\,\,\, 2\varepsilon_{12}=\varepsilon_{3}
=u_{,\,y} + v_{,\,x}\,.
\]
The solvability of this system with regard to functions $u$ and $v$ is
provided for by a traditional strain compatibility equation [9, 10]
\[
 \varepsilon_{11,\,yy} + \varepsilon_{22,\,xx} - \,
2\varepsilon_{12,\,xy} = 0\,.
\]

As long as $C_{12}= C_{11} \!- 2 C_{33}$, matrix $A$ has got only three
independent components. Let us introduce three kinetic parameters
$\lambda _0\,,\, \mu _0\,,\, \mu$\,,  so that $C_{11}= \lambda _0 +
\mu$\,, $C_{12}=\lambda _0-\mu$\,, $C_{13}=\mu _0$\,, $C_{33}= \mu$. Then
relations (2.1) can be put in a tensor form
\begin{equation}
T = I\,\lambda _0\,\mathrm{div}\,w + 2M\,\gamma \,,
\end{equation}
\[
T = \left( {{\begin{array}{*{20}c}
   {\sigma_{11}} \hfill & {\sigma_{12}} \hfill \\
   {\sigma_{12}} \hfill & {\sigma_{22}} \hfill \\
\end{array} }} \right) \,, \quad
M = \left( {{\begin{array}{*{20}c}
   \,\,\,\mu \hfill &  {\mu _0 } \hfill \\
 \!\!{ - \mu _0 } \hfill & \,\mu \hfill \\
\end{array} }} \right) \,, \quad
 2\gamma \, = \left( {{\begin{array}{*{20}c}
 {u_{,\,x} - v_{,\,y} } \hfill & {u_{,\,y} + v_{,\,x} } \hfill \\
 {v_{,\,x} + u_{,\,y} } \hfill & {v_{,\,y} - u_{,\,x} } \hfill \\
\end{array} }} \right) \,,
\]
where $I$ is a unit tensor, $T$ is a stress tensor, $M$ is the
kinetic parameters tensor, and $\gamma$ is a strain
tensor-deviator. Formula (2.2) reveals physical sense of
parameters $\mu$ and $\mu_0$. These are shear modules of an
elastic body under the conditions of plane deformation with one
admissible rotation operator.

Constitutive equations of type (2.2) were obtained earlier by V.O.
Bytev [6] for viscous stresses. As opposed to traditional
equations, they contain three kinetic parameters: $\lambda _0\,,\,
\mu_0\,,\, \mu$, where $\lambda_0>0$\,, $\mu>0$\,, and $\mu_0$ may
have any real value. It is also worth mentioning that relation
$\mu_0/\mu$ defines the measure of noncoaxiality of stress and
strain tensors-deviators [6], and the substitution $\lambda_0=
\lambda+ \mu$\,, $\mu_0=0$\, will turn (2.2) into a traditional
system of constitutive equations with parameters $\lambda$ and
$\mu$ [9,~10].

Relations opposite to (2.2) are represented in the following way:
\begin{equation}
\begin{array}{l}
 u_{,\,x} = {\textstyle{1 \over 4}}\left[ {\lambda _0^{ - 1} \left(
{\sigma _{11} + \sigma _{22} } \right) + \mu \kappa _0^{ - 2} \left(
{\sigma _{11} - \sigma _{22} } \right) - 2\mu _0 \kappa
_0^{ - 2} \sigma _{12} } \right] , \\
 v_{,\,y} = {\textstyle{1 \over 4}}\left[ {\lambda _0^{ - 1} \left(
{\sigma _{11} + \sigma _{22} } \right) - \mu \kappa _0^{ - 2} \left(
{\sigma _{11} - \sigma _{22} } \right) + 2\mu _0 \kappa _0^{ - 2}
\sigma _{12} } \right] , \\
 u_{,\,y} + v_{,\,x} = {\textstyle{1 \over 2}}
\, \kappa_0^{-2} \left[ { \mu_0 \left( \sigma _{11} - \sigma_{22} \right)
+ 2\mu \, \sigma _{12} } \right] ,
\end{array}
\end{equation}
where \ $\kappa _0^2 = \mu ^2 + \mu _0^2$\,.

 Let us take the homogeneous equations of plane problem statics
\[
\sigma_{11,\,x}+\sigma_{12,\,y}=0\,, \quad
\sigma_{12,\,x}+\sigma_{22,\,y}=0\,
\]
 and introduce Airy function $U$ and function $Q$:
 $Q = \sigma_{11} + \sigma_{22} = \Delta U, \ \sigma_{11} =
 U_{,\,yy}= Q - U_{,\,xx}\,, \ \sigma_{22}= U_{,\,xx}= Q - U_{,\,yy} \
 \sigma_{12}= - U_{,\,xy}\,.$
 the first two equations (2.3) shall be transformed as follows:
\begin{equation}
\begin{array}{l}
 u_{,\,x} = {\textstyle{1 \over 4}}\left( {\lambda _0^{ - 1} + \mu
\kappa _0^{ - 2} } \right)Q - {\textstyle{1 \over 2}} \,\mu \kappa _0^{ -
2} U_{,\,xx} + {\textstyle{1 \over 2}} \,\mu _0 \kappa _0^{ -
2} U_{,\,xy} \ , \\
 v_{,\,y} = {\textstyle{1 \over 4}}\left( {\lambda _0^{ - 1} + \mu
\kappa _0^{ - 2} } \right)Q - {\textstyle{1 \over 2}} \,\mu \kappa _0^{ -
2} U_{,\,yy} - {\textstyle{1 \over 2}} \,\mu _0 \kappa _0^{ - 2}
U_{,\,xy} \ .
\end{array}
\end{equation}
Now following the same pattern as in [10] we will find that $\Delta Q =
0$. Therefore we may conclude that $Q$ is a harmonic function. Let us
take $R$ as a conjugate harmonic function in relation to $Q$:
\[
\frac{\partial Q}{\partial x} = \frac{\partial R}{\partial y} \,\,, \quad
\frac{\partial Q}{\partial y} = - \frac{\partial R}{\partial x}
\]
and $Q + i R = f(z)$, where $f$ is an analytic function of argument $z =
x + i y$.

Now let us introduce an analytic function $\varphi(z) = p + iq =
\textstyle{1\over 4}\int {f\left( z \right)}dz$\,. In compliance to the
Cauchy-Riemann conditions the following equations are formed [10]
\[
 \frac{\partial p}{\partial x} = \frac{\partial q}{\partial y} =
 {\textstyle{1 \over 4}}\, Q\,, \quad
 \frac{\partial p}{\partial y} =
-\frac{\partial q}{\partial x} = {\textstyle{1 \over 4}}\, R\,.
\]
With the help of them, equations (2.4) are transformed as follows:
\[
    u_{,\,x} = \left( {\lambda _0^{ - 1} + \mu\kappa _0^{ - 2} }
\right)p_{,\,x} - {\textstyle{1 \over 2}}\,\mu\kappa _0^{ - 2}
  U_{,\,xx} + {\textstyle{1 \over 2}} \,\mu _0\kappa _0^{ - 2}U_{,\,xy}\,,
\]
\[
    v_{,\,y} = \left( {\lambda _0^{ - 1} + \mu\kappa _0^{ - 2} }
\right)q_{,\,y} - {\textstyle{1 \over 2}}\,\mu\kappa _0^{ - 2}
  U_{,\,yy} - {\textstyle{1 \over 2}} \,\mu _0\kappa _0^{ - 2}U_{,\,xy}\,.
\]
The following representations for displacement vector components are
received after integration:
\begin{equation}
\begin{array}{l}
 u = \left( {\lambda _0^{ - 1} + \mu \kappa _0^{ - 2}} \right)p -
 {\textstyle{1 \over 2}}\,\mu   \kappa _0^{ - 2} U_{,\,x} +
 {\textstyle{1 \over 2}}\,\mu _0\kappa _0^{ - 2} U_{,\,y} + f_1 (y)\,, \\
 v = \left( {\lambda _0^{ - 1} + \mu \kappa _0^{ - 2}} \right)q -
 {\textstyle{1 \over 2}}\,\mu   \kappa _0^{ - 2} U_{,\,y} -
 {\textstyle{1 \over 2}}\,\mu _0\kappa _0^{ - 2} U_{,\,x} + f_2 (x)\,. \\
\end{array}
\end{equation}

The last equation in the system (2.3) is transformed as follows:
\begin{equation}
 u_{,\,y} + v_{,\,x} = {\textstyle{1 \over 2}}\,\mu _0\kappa _0^{ - 2}
 \left( {U_{,\,yy} - U_{,\,xx}} \right) - \mu\kappa _0^{ - 2} U_{,\,xy}\,.
\end{equation}
After calculation of corresponding derived functions (2.5) --
\[
 u_{,\,y} = \left( {\lambda _0^{ - 1} + \mu\kappa _0^{ - 2}} \right)
 p_{,\,y} - {\textstyle{1 \over 2}}\,\mu\kappa _0^{ - 2} U_{,\,xy} +
 {\textstyle{1 \over 2}}\,\mu _0\kappa _0^{- 2} U_{,\,yy} + f_{1,\,y}\,,
\]
\[
 v_{,\,x} = \left( {\lambda _0^{ - 1} + \mu\kappa _0^{ - 2}} \right)
 q_{,\,x} - {\textstyle{1 \over 2}}\,\mu\kappa _0^{ - 2} U_{,\,xy} -
 {\textstyle{1 \over 2}}\,\mu _0\kappa _0^{- 2} U_{,\,xx} + f_{2,\,x}\,,
\]
we get their sum as compared to (2.6):
\[
 u_{,\,y} + v_{,\,x} = {\textstyle{1 \over 2}}\,\mu _0\kappa _0^{ - 2}
 \left( {U_{,\,yy} - U_{,\,xx}} \right) - \mu\kappa _0^{ - 2} U_{,\,xy} =
\]
\[
(\lambda _0^{ - 1} + \mu \kappa _0^{ - 2} )\left( {p_{,\,y} + q_{,\,x}}
\right) - \mu \kappa _0^{ - 2} U_{,\,xy} + {\textstyle{1 \over 2}}\,\mu
_0 \kappa _0^{ - 2} \left( {U_{,\,yy} - U_{,\,xx}} \right) + f_{1,\,y} +
f_{2,\,x} \,.
\]
With account of relation $q_{,\,x} = - p_{,\,y}$\, we get the following
equations $u_{,\,y} + v_{,\,x} = f_{1,\,y} + f_{2,\,x} = 0$\,, which
means that
\begin{equation}
 f_1 \left( y \right) = c \left( {- \varepsilon y + \alpha _1} \right)\,,
 \ f_2 \left( x \right) = c \left( {\varepsilon x + \alpha _2} \right)\,,
\end{equation}
where $c ,\,\,\alpha _1 ,\,\,\alpha _2 ,\,\,\varepsilon$ are constant
values.

Now lets us go back to (2.5). Taking into account that additional
summands of the type (2.7), define rigid translation eliminated by
transfer into a new system of coordinates, we get simpler representations
for displacement vector components:
\begin{equation}
\begin{array}{l}
 u = \left( {\lambda _0^{ - 1} + \mu \kappa _0^{ - 2}} \right)p -
 {\textstyle{1 \over 2}} \, \mu \kappa _0^{ - 2} U_{,\,x} +
 {\textstyle{1 \over 2}}\,\mu _0\kappa _0^{ - 2} U_{,\,y}\,, \\
 v = \left( {\lambda _0^{ - 1} + \mu \kappa _0^{ - 2}} \right)q -
 {\textstyle{1 \over 2}} \, \mu \kappa _0^{ - 2} U_{,\,y} -
 {\textstyle{1 \over 2}}\,\mu _0\kappa _0^{ - 2} U_{,\,x}\,. \\
 \end{array}
\end{equation}
where $p = p\left( {x,y} \right), \ q = q\left( {x,y}\right)$.

After introduction of a complex relations definition (2.8)
\[
u + iv = \left( {\lambda _0^{ - 1} + \mu \kappa _0^{ - 2}} \right) \left(
{p + iq} \right) - {\textstyle{1 \over 2}}\,\kappa \kappa _0^{ - 2}
\left( {U_{,\,x} + iU_{,\,y}} \right),
\]
symbol $\kappa = \mu + i\mu _0$ and taking into account that $p + \!iq =
\varphi \left( z \right)$, and $U$ is a biharmonic function, we shall get
the following complex representation for a displacement vector:
\begin{equation}
2\kappa _0^2 \left( {u + iv} \right) = \left( {2\lambda _0^{ - 1} \kappa
_0^2 + \overline{\kappa}} \right) \varphi \left( z \right) - \kappa
\,z\overline {{\varphi }'\left( z \right)} - \kappa \,\overline {\psi
\left( z \right)} \,,
\end{equation}
which is a extension of traditional Love formulas [8,~10] (prime
means derivative with respect to $z$).

As for complex representation of stress components with the help of similar
functions $\phi $ and $\psi $ used for the representation of a biharmonic
Airy function, it is no way different from the traditional one [10]:
\[
\sigma _{11} + \sigma _{22} = 2\left[ {{\varphi}'\left( z \right) +
\overline {{\varphi}'\left( z \right)}} \right] = 4\textstyle{Re} \left[
{{\varphi}'\left( z \right)} \right], \ \sigma _{22} - \sigma _{11} +
i\sigma _{12} = 2\left[{\overline{z} {\varphi}''\left( z \right) +
{\psi}''\left( z \right)} \right].
\]

Boundary condition in the second basic problem of the elasticity theory
regarding determination of elastic equilibrium on condition of the given
boundary displacements shall be represented as follows:
\[
\left( {2\lambda _0^{ - 1} \kappa _0^2 + \overline{\kappa}} \right)
\varphi \left( z \right) - \kappa \,z\overline {{\varphi}' \left( z
\right)} - \kappa \,\overline {\psi \left( z \right)} = 2\kappa _0^2
\left( {q_1 + iq_2 } \right),
\]
where $q_{1}$ and $q_{2}$ are set displacements of boundary points.

Let us assume that two complex planes $Z$ and $G$ and conformal mapping
$z = \omega \left( \zeta \right)$ of area $S \subset Z$ to area $\Sigma
\subset G$ are given. Now let us introduce polar coordinates
$\left({r,\,\theta}\right)$ at G plane, so that $\omega (\zeta ) = \zeta
= r e^{i\theta }$. Then any vector $\left( {w_x \,,\, w_y } \right)$ is
transformed in compliance to the following formula: $w_r + \,iw_\theta =
e^{ - i\theta }\left( {w_x + \,iw_y } \right)$.With the help of this we
can get polar representation of displacement vector
$w=(u_r\,,\,u_\theta)$ from (2.9):
\begin{equation}
2\kappa _0^2 \left[ {u_r + iu_\theta} \right] = e^{- i\theta} \left(
{\left( {2\lambda _0^{ - 1} \kappa _0^2 + \overline{\kappa}} \right)
\varphi \left( \zeta \right) - \kappa \,\zeta \,\overline
{{\varphi}'\left( \zeta \right)} - \kappa \,\overline {\psi \left( \zeta
\right)}} \right).
\end{equation}
Let us proceed to the following problems in order to demonstrate new
effects of plane deformation of asymmetrically-elastic plates.
  \vskip 2mm

{\bf Circular Washer under Uniform Pressure}. In this problem as in
traditional elasticity we see the following [10]:
\[
\varphi \left( z \right) = -\,{\textstyle{1 \over 2}} \,p z\,, \,\,\,
\psi \left( z \right) = 0\,, \,\,\, \sigma _{rr} = - p\,, \,\,\, \sigma
_{\theta\theta} = - p\,, \,\,\, \sigma _{r\theta} = 0\,.
\]
As opposed to the traditional solution, displacement vector (2.10) has
got not only radial but also angular component vanishing when
$\mu_{0}=0$:
\[
u_r = -\, {\textstyle{1 \over 2}}\,\lambda _0^{ - 1} pr, \,\,\,\,
u_\theta= {\textstyle{1 \over 2}}\,\mu _0 \kappa _0^{ - 2} pr.
\]
  \vskip 2mm

{\bf Biaxial Tension of a Plate with a Circular Hole.} In this case we
get the following [10]:
\[
\varphi \left( z \right) = \textstyle{1 \over 2}\,p z\,,\,\,\,\,
   \psi \left( z \right) = - p R^2 z^{ - 1},
\]
\[
\sigma _{rr} = p \left( {1 - \frac{R^2}{r^2}} \right), \,\,\,\, \sigma
_{\theta \theta } = p \left( {1 + \frac{R^2}{r^2}} \right), \,\,\,\,
\sigma _{r\theta } = 0
\]
(R is the hole radius). Relevant (2.10) displacement vector
components are represented by the formulas
\begin{equation}
u_r = {\textstyle{1 \over 2}} \, pR  \left( {\lambda _0^{ - 1}
{\frac{r}{R}} + \mu \kappa _0^{ - 2} {\frac{R}{r}}} \right)\,,\,
\,\,\,u_\theta = {\textstyle{1 \over 2}}\,pR {\mu _0 \kappa _0^{- 2}
\left( { {\frac{R}{r}} - {\frac{r}{R}} } \right)}\,,
\end{equation}
and their values at boundary $r = R$ -
\begin{equation}
\left. {u_r} \right|_{r = R} = {\textstyle{1 \over 2}} \,{pR\left( {
\lambda _0^{ - 1} + \mu \kappa _0^{ - 2} } \right)}, \,\,\,\, \left.
{u_\theta} \right|_{r = R} = 0\,.
\end{equation}

Taking into account that $\kappa _0^2 = \mu ^2 + \mu _0^2$ and
representing parameter $\lambda _{0}$ in the form of the sum
$\lambda_{0}=\lambda +\mu $, we pass to the limit in (2.11) and (2.12)
with $\mu \to 0$:
\[
u_r \to \frac{p\,r}{2\lambda } \ , \,\,\,\, u_\theta \to \frac{pR}{2\mu
_0 }\left( {\frac{R}{r} - \frac{r}{R}} \right), \,\,\,\, \left. {u_r}
\right|_{r = R} \to \frac{pR}{2\lambda} \ , \,\,\,\, \left. {u_\theta}
\right|_{r = R} = 0\,.
\]
If $\mu _0 = 0$, formulas (2.11) take a traditional form [10]
\[
u_r^{cl} = {\textstyle{1 \over 2}}\, pR\left( {\lambda _0^{ - 1}
\frac{r}{R} + \mu ^{ - 1}\frac{R}{r}}\, \right) , \,\,\,\, u_\theta ^{cl}
\equiv 0\,.
\]
When $\mu \to 0$, passage to the limit here does not have any physical
sense.
  \vskip 2mm

{\bf Uniaxial Tension of a Plate with a Circular Hole}. Let us assume
that the contour of the hole is free from external stress and at infinity
$\sigma _{11}^\infty = p\,,\,\,\sigma _{22}^\infty = 0\,,\,\,\sigma
_{12}^\infty = 0$\,. It means that tension is present along $Ox$ axis,
and tensile stress at infinity being a constant value $p$\,. In this case
functions $\varphi(z)$, $\psi(z)$ and stress tensor components shall be
defined with the help of the following equations [10]:
\[
\varphi \left( z \right) = {\textstyle{1 \over 4}}\, pR \left(
{\frac{z}{R} + 2\frac{R}{z}} \right)\,, \,\,\,\, \psi \left( z \right) =
-\, {\textstyle{1 \over 2}}\, pR \left( {\frac{z}{R} + \frac{R}{z} -
\frac{R^3}{z^3}} \right)\,,
\]
\[
\sigma _{rr} = \frac{p}{2}\left[ {1 - \frac{R^2}{r^2} + \left( {1 -
4\frac{R^2}{r^2} + 3\frac{R^4}{r^4}} \right)\cos 2\theta } \right]\,,
\]
\[
\sigma _{\theta \theta } = \frac{p}{2}\left[ {1 + \frac{R^2}{r^2} -
\left( {1 + 3\frac{R^4}{r^4}} \right)\cos 2\theta } \right]\,, \,\,\,\,
\sigma _{r\theta } = - \frac{p}{2}\left( {1 + 2\frac{R^2}{r^2} -
3\frac{R^4}{r^4}} \right)\sin 2\theta\,.
\]
Relevant (2.11) displacement vector components are as follows:
\[
u_r = {\textstyle{1 \over 4}}\, pR\left( {\lambda _0^{ - 1} \frac{r}{R} +
\mu \kappa _0^{ - 2} \frac{R}{r}} \right) + {\textstyle{1 \over 4}}\, \mu
_0 \kappa _0^{ - 2}\, pR\left( {\frac{r}{R} - 2\frac{R}{r} +
\frac{R^3}{r^3}} \right)\sin 2\theta \ +
\]
\begin{equation}
{\textstyle{1 \over 4}}\, pR\left[ {2\lambda _0^{ - 1} \frac{R}{r} + \mu
\kappa _0^{ - 2} \left( {\frac{r}{R} + 2\frac{R}{r} - \frac{R^3}{r^3}}
\right)} \right]\cos 2\theta\,,
\end{equation}
\[
u_\theta = {\textstyle{1 \over 4}}\, \mu _0 \kappa _0^{ - 2} pR\left(
{\frac{R}{r} - \frac{r}{R}} \right) + {\textstyle{1 \over 4}}\, \mu _0
\kappa _0^{ - 2} pR\left( {\frac{r}{R} - \frac{R^3}{r^3}} \right)\cos
2\theta \ -
\]
\begin{equation}
{\textstyle{1 \over 4}}\, pR\left[ {2\lambda _0^{ - 1} \frac{R}{r} + \mu
\kappa _0^{ - 2} \left( {\frac{r}{R} + \frac{R^3}{r^3}} \right)}
\right]\sin 2\theta\,.
\end{equation}

Assuming that $\mu _0 = 0$, we get traditional formulas of this problem
[10]:
\[
u_r^{cl} = {\textstyle{1 \over 4}}\, pR\left( {\lambda _0^{ - 1}
\frac{r}{R} + \mu ^{ - 1}\frac{R}{r}} \right) + {\textstyle{1 \over 4}}\,
pR\left[ {2\lambda _0^{ - 1} \frac{R}{r} + \mu ^{ - 1} \left(
{\frac{r}{R} + 2\frac{R}{r} - \frac{R^3}{r^3}} \right)} \right]Ьcos
2\theta\,,
\]
\[
u_\theta ^{cl} = -\, {\textstyle{1 \over 4}}\, pR\left[ {2\lambda _0^{ -
1} \frac{R}{r} + \mu ^{ - 1}\left( {\frac{r}{R} + \frac{R^3}{r^3}}
\right)} \right]\sin 2\theta\,.
\]

Now let us calculate displacement values at the contour $r = R$ with the
help of (2.13) and (2.14):
\[
 \left. {u_r } \right|_{r = R} = \frac{pR\,\left( {\mu _0^2 + 2\mu
^2 + \lambda \mu } \right)}{4\left( {\lambda + \mu } \right)\left( {\mu
_0^2 + \mu ^2} \right)}\left( {1 + 2\cos 2\theta } \right),
 \,\,\, \left. {u_\theta } \right|_{r = R} =
- \frac{pR\,\left( {\mu _0^2 + 2\mu ^2 + \lambda \mu } \right)}{2\left(
{\lambda + \mu } \right)\left( {\mu _0^2 + \mu ^2} \right)}\sin 2\theta .
\]
Passage to the limit with $\mu \quad \to$ 0 will give final values of
boundary displacements:
\[
\left. {u_r } \right|_{r = R} \to \frac{pR}{4\lambda }\left( {1 + 2\cos
2\theta } \right), \quad \left. {u_\theta} \right|_{r = R} \to -
\frac{pR}{2\lambda }\sin 2\theta .
\]

Traditional formulas give the following values
\[
\left. {u_r^{cl} } \right|_{r = R} = \frac{pR\left( {\lambda + 2\mu }
\right)}{4\mu \left( {\lambda + \mu } \right)}\left( {1 + 2\cos 2\theta }
\right), \quad \left. {u_\theta ^{cl} } \right|_{r = R} = -
\frac{pR\left( {\lambda + 2\mu } \right)}{2\mu \left( {\lambda + \mu }
\right)}\sin 2\theta ,
\]
having no physical sense when $\mu \to 0$.
  \vskip 4mm

{\bf Conclusion}. A non-traditional version of the elasticity theory
suggested herein contains additional kinetic parameters and requires
special experiments for their evaluation. Thus, it provides a scientific
basis and new opportunities for experimental analysis. The authors hope
that the abandonment of a traditional condition of elasticity tensor
symmetry shall considerably expand the opportunities of both linear and
nonlinear elasticity theories. Asymmetric non-polar medium theory can be
applicable to modeling of anomalies connected to media and materials
microstructure.

\newpage
\begin{center}
BIBLIOGRAPHY
\end{center}

\begin{enumerate}
\item C. Truesdell. A first course in rational continuum mechanics.
 Baltimore: The Johns Hopkins University, 1972.
\item L.V. Ovsyannikov. Group analysis of differential equations.
 Moskow: Nauka.
\item N.Kh. Ibragimov. Groups of transformations in mathematical physics.
 Moskow: Nauka, 1983.
\item B.D. Annin, V.O. Bytev, S.I. Senashev. Group properties of equations
 of elasticity and plasticity. Novosibirsk: Nauka, 1985.
\item V.O. Bytev. Building of Mathematical Models of continuum media
 on the basis of invariance principle//Acta Appl. Math. / Kluwer Acad. Pupl.,
 Netherlands, 1989. Vol.16. P.117-142.
\item V.K. Andreev, V.V. Bublik, V.O. Bytev. Symmetry of non-traditional
 models of hydrodynamics. Novosibirsk: Nauka, 2003.
\item F.R. Gantmakher. Theory of matrices/ 5th ed. Moskow: Physmatlit, 2004.
\item A.E.H. Love. The mathematical theory of elasticity/ 4th ed. Cambridge, 1927.
\item S.P. Timoshenko, J.N. Goodier. Theory of elasticity/ 3rd ed. N.Y.:
 McGraw-Hill, 1970.
\item N.I. Muskhelishvili. Some basic problems of the mathematical theory
 of elasticity. Moskow: Nauka, 1966.
\end{enumerate}
  \bigskip

V.O. Bytev, Tyumen, e-mail: vbytev@utmn.ru
  \bigskip

L.I. Shkutin, Krasnoyarsk, e-mail: shkutin@icm.krasn.ru.

\end{document}